\documentstyle[12pt,psfig]{article}
\def\lf{\leaders\hbox to 1em{\hss.\hss}\hfill}
\def\21{$SU(2) \ot U(1)$}
\def\321{$SU(3) \ot SU(2) \ot U(1)$}
\def\eq#1{{eq. (\ref{#1})}}
\def\Eq#1{{Eq. (\ref{#1})}}

\def\VEV#1{\left\langle #1\right\rangle}

\def\lsim{\raise0.3ex\hbox{$\;<$\kern-0.75em\raise-1.1ex
\hbox{$\sim\;$}}}
\def\gsim{\raise0.3ex\hbox{$\;>$\kern-0.75em\raise-1.1ex
\hbox{$\sim\;$}}}

\def\beq{\begin{equation}}
\def\eeq{\end{equation}}
\def\bef{\begin{figure}}
\def\eef{\end{figure}}
\def\bet{\begin{table}}
\def\eet{\end{table}}
\def\bea{\begin{eqnarray}}
\def\eea{\end{eqnarray}}
\def\ba{\begin{array}}
\def\ea{\end{array}}
\def\bi{\begin{itemize}}
\def\ei{\end{itemize}}
\def\ben{\begin{enumerate}}
\def\een{\end{enumerate}}

\def\ot{\otimes}

\def\np#1#2#3{    { Nucl. Phys. }{\bf #1}, #2 (19#3)}
\def\pl#1#2#3{    { Phys. Lett. }{\bf #1}, #2 (19#3)}
\def\pr#1#2#3{    { Phys. Rev. }{\bf #1}, #2 (19#3)}

\def\prl#1#2#3{   { Phys. Rev. Lett. }{\bf #1}, #2 (19#3)}

\def\ptp#1#2#3{   { Prog. Theor. Phys. }{\bf #1}, #2 (19#3)}

\def\Zp#1#2#3{    { Z. Physik }{\bf #1}, #2 (19#3)}

\def\n.c.#1#2#3{  { Nuovo Cim. }{\bf #1}, #2 (19#3)}
\def\r.n.c.#1#2#3{{ Riv. del Nuovo Cim. }{\bf #1}, #2 (19#3)}
\def\sjnp#1#2#3{  { Sov. J. Nucl. Phys. }{\bf #1}, #2 (19#3)}

\def\ppnp#1#2#3{  { Prog. Part. Nucl. Phys. }{\bf #1}, #2 (19#3)}

\renewcommand{\thefootnote}{\fnsymbol{footnote}}
\def\dfrac#1#2{{\displaystyle\frac{#1}{#2}}}

\newcommand{\mat}[4]{\left(\begin{array}{cc}{#1}&{#2}\\{#3}&{#4}\end{array}
\right)}
\newcommand{\matr}[9]{\left(\begin{array}{ccc}{#1}&{#2}&{#3}\\{#4}&{#5}&{#6}\\
{#7}&{#8}&{#9}\end{array}\right)}           

\begin{document}
\thispagestyle{empty}
\begin{titlepage}
\begin{center}
\rightline{hep-ph/9605301}
\hfill FTUV/96-25\\
\hfill IFIC/96-29\\
\hfill \today \\
\vskip 0.2cm
{\Large \bf Resonant Conversion of Massless Neutrinos in Supernovae\\
}
\vskip 0.2cm
{H.  Nunokawa}$^1$ 
\footnote{
E-mail: nunokawa@flamenco.ific.uv.es},
Y.-Z. Qian$^2$ 
\footnote{
E-mail: yzqian@citnp.caltech.edu},
A. Rossi$^1$
\footnote{
E-mail: rossi@evalvx.ific.uv.es, rossi@ferrara.infn.it},
and 
J. W. F. Valle$^1$
\footnote{
E-mail: valle@flamenco.ific.uv.es}\\
{\sl $^1$Instituto de F\'{\i}sica Corpuscular - C.S.I.C.\\
Departament de F\'{\i}sica Te\`orica, Universitat de Val\`encia\\
46100 Burjassot, Val\`encia, SPAIN\\
URL http://neutrinos.uv.es}\\
\vskip .2cm
{\sl $^2$Physics Department, 161-33\\
California Institute of Technology\\
Pasadena, CA 91125, USA} 
\vskip .2cm
{\bf Abstract}
\end{center}
\begin{quotation}
It has been noted for a long time that, in some circumstances,
{\sl massless} neutrinos may be {\sl mixed} in the leptonic charged 
current. Conventional neutrino oscillation searches in vacuum are 
insensitive to this mixing. We discuss the effects of resonant 
massless-neutrino conversions in the dense medium of a supernova. 
In particular, we show how the detected $\bar\nu_e$ energy spectra 
from SN1987a and the supernova $r$-process nucleosynthesis  
may be used to provide very stringent constraints on the mixing of
{\sl massless} neutrinos. 

\end{quotation}
\end{titlepage}

\renewcommand{\thefootnote}{\arabic{footnote}}
\setcounter{footnote}{0}
\section{Introduction}
\ 

In the original scenario developed by Mikheyev and Smirnov \cite{MS} 
the resonant neutrino conversion in matter requires non-degenerate 
neutrino masses and non-vanishing mixing angles in vacuum. 
In the basis of two neutrino flavour eigenstates, the evolution 
Hamiltonian describing the neutrino propagation in matter is given by 
\begin{eqnarray}
\label{hamil}
 {\cal H} & \!  
= & \! \mat{H_e}{H_{e\alpha}}{H_{e\alpha}}{H_{\alpha}}, \,\,\,\,\,
\,\,\,\, \alpha=\mu\ (\tau), \\
  H_e & \! = &  \! V_e - \frac{\delta m^2}{4E_\nu} \cos2 \theta, \,\,\,
H_\alpha \!= \!V_\alpha + \frac{\delta m^2}{4E_\nu} \cos2 \theta, \nonumber \\
H_{e\alpha}& \!= & \!H_{\alpha e} = \frac{\delta m^2}{4E_\nu} \sin2 \theta, 
\nonumber
\end{eqnarray}
where $V_e$ and $V_\alpha$ are the well-known Wolfenstein {\sl diagonal} 
matter potentials arising from coherent neutrino scatterings off 
matter particles \cite{W}. In Eq. (1), $E_\nu$ is the neutrino energy, and
$\delta m^2$ 
and $\theta$ are the neutrino mass-squared difference and the mixing angle
in vacuum, respectively. 
One can see that the {\sl effective} mixing in matter between $\nu_e$ 
and $\nu_\alpha$ states is induced by the ``vacuum"  term 
$(\delta m^2/4E_\nu) \sin 2 \theta$. 

It has been noticed for a long time that the presence of \21 isosinglet 
neutral heavy leptons \cite{fae} in general leads to flavour-changing 
neutral-current (FCNC) interactions of neutrinos \cite{2227}. As a result,
there can be non-trivial leptonic mixing (and CP violation) \cite{NHL} 
involving the conventional isodoublet neutrinos even in models where 
these neutrinos remain strictly massless, as in the Standard Model, 
due to an exactly conserved lepton number \cite{MV}. The non-vanishing 
massless-neutrino mixing angles arise due to the presence of extra 
heavy gauge singlet neutral states.
In this scenario the interaction of massless neutrinos with matter 
constituents gives rise to a non-trivial neutrino evolution 
Hamiltonian, analogous to \Eq{hamil} \cite{valle,LL}, which can
mix the neutrino identities \cite{valle,LL}. This Hamiltonian is 
characterized by a new type of weak potentials whose  diagonal 
and off-diagonal matrix elements will be discussed below.

The implications of both {\sl standard} and {\sl non-standard}  
neutrino interactions for the neutrino propagation in dense media have 
been extensively studied \cite{KP,BPW}. In particular, the birth of neutrino 
astronomy, with the detection of neutrinos from the Sun \cite{BAH} and 
SN1987a \cite{KA,IMB}, has offered the 
opportunity to probe various neutrino properties, such as neutrino masses 
and mixings, neutrino lifetimes, neutrino magnetic moments, and
generically, any {\sl non-standard} interactions of neutrinos.

In this paper we focus on the particular scenario 
of massless-neutrino mixing suggested in Ref. 
\cite{valle}. This scenario can be relevant only for the neutrino  
propagation in strongly-neutronized media. Such media 
exist perhaps only in supernovae.  
We show how to probe the mixing in the light neutrino sector by 
considering two different aspects of the supernova process. 
We examine how the
massless-neutrino conversion of the type
 $\bar{\nu}_e \leftrightarrow \bar{\nu}_\alpha$ 
can affect the detected $\bar{\nu}_e$ energy spectra \cite{old,SSB,JNR} 
from SN1987a. 
We also consider the implications of such conversions for 
the supernova $r$-process nucleosynthesis, following the same 
lines of reasoning adopted in  Ref. \cite{QFMMW}.
Rather stringent limits on the mixing of massless neutrinos may
be derived from both considerations. These limits are very remarkable
because the mixing of massless neutrinos cannot be sharply constrained 
through neutrino oscillation 
searches. Being strictly massless, these neutrinos cannot 
develop any phase difference {\sl in vacuum}, and as a result,
neutrino oscillations can not occur. 

In Sec. 2  we give a quick reminder on the theoretical framework of Ref.
\cite{valle}. In Sec. 3 we present the general features of the 
resonant massless-neutrino conversion in matter. Sec. 4 discusses
resonant conversions of massless-neutrinos in supernovae and the implications 
of such conversions for supernova neutrino detection and $r$-process
nucleosynthesis.
We summarize our results and conclude in Sec. 5. 
 
\section{The Massless and Mixed Neutrino Model}
\

In the Standard Model the absence of right-handed neutrino states 
naturally implies that neutrinos stay massless to all orders of 
perturbation even after the gauge symmetry breaking and 
there are no Cabibbo-Kobayashi-Maskawa-like \cite{CKM} mixing matrices 
in the weak leptonic charged current. In this case the total lepton 
number $L$ comes out as an {\sl accidental} symmetry
\cite{TH} 
due to the gauge structure and renormalizability of the theory. 

On the other hand, any number of gauge singlet neutral leptons can be 
introduced since they do not carry triangle anomaly \cite{2227}. These 
extra states can arise in left-right symmetric, grand-unified or 
superstring-inspired models \cite{MV,WW,Witten,beyond}. 
In this case the lepton number is no longer an {\sl accidental} 
symmetry and it may be imposed {\sl by hand}.
The simplest such scheme \cite{NHL,MV,WW} contains three two-component
gauge singlet neutral leptons $S$ added to the three right-handed 
neutrino components $\nu^c$ present in SO(10). For definiteness
we consider this model at the \21 level. The assumed conservation 
of lepton number leads to a neutral mass matrix with the following 
texture in the basis ($\nu, \nu^c, S$): 
\beq
\matr{0}{D}{0}
{D^T}{0}{M}{0}{M^T}{0},
\label{matrix}
\eeq
where the Dirac matrix $D$ describes the coupling between the 
weak doublet $\nu$ and the singlet $\nu^c$, and where the other Dirac 
matrix $M$ connects the singlet states $\nu^c$ and $S$. 
It is easy to see that, as expected, the three conventional 
neutrinos remain massless, while the other six neutral 2-component
leptons combine into three heavy Dirac fermions \cite{NHL,WW}. 

The phenomenological implications of this picture are manifest 
when considering  the resulting charged-current (CC) Lagrangian 
in the massless-neutrino sector:
\beq
\label{CC}
{\cal L}_{\rm CC}= \frac{\mbox{i} g}{\sqrt{2}} W_\mu \bar{e}_{a L} 
\gamma_\mu  K_{a i} \nu_{i L} + \mbox{h.c.}, \,\,\,\,\,\,\,a=e,\mu,\tau, 
\,\,\,\,i=1,2,3,
\eeq
where the mixing matrix $K$ is not unitary, since it is a sub-matrix of 
the full rectangular matrix including also the heavy states \cite{2227}. 
Therefore, the non-diagonal elements of the matrix $K$ cannot be rotated 
away through a redefinition of the massless-neutrino fields. In this way
a non-vanishing mixing arises among the massless neutrinos. 
The corresponding form of the neutral-current (NC) Lagrangian for
the massless-neutrino sector is 
\beq
\label{NC}
{\cal L}_{\rm NC}= \frac{\mbox{i} g}{2 \cos \theta_{W}} Z_\mu P_{ij} 
\bar{\nu}_{i L} 
\gamma_\mu  \nu_{j L}\:,
\eeq
where $P= K^{\dagger} K$. 
Unlike in the Standard Model, the matrix $P$ is diagonal but 
generation-dependent, signalling the violation of weak universality. 

For definiteness, we later on use an explicit parametrization 
of the matrix $K$, confining ourselves to the case of two (massless) 
neutrinos $\nu$. 
We may write the mixing matrix $K$ as \cite{2227,valle}
\begin{equation}
K = R\,{\cal N},
\end{equation}
where $R$ is a $2 \times 2$ rotation matrix,
\begin{equation}
R= \left(\matrix{\cos \theta &\sin \theta \cr 
                - \sin \theta &\cos \theta } \right),
\label{mixing}
\end{equation}
and where the diagonal matrix, 
\begin{equation}
{\cal N} =  \mat{{\cal N}_1} {0} 
                  {0} {{\cal N}_{2,3}},
\label{nonuniversal}
\end{equation}
describes the effective non-orthogonality of the two neutrino
flavours, i.e., \\ 
$\langle\nu_e|\nu_{\mu,\tau}\rangle \equiv - \sin\theta  \cos\theta 
({\cal N}_1^2 - {\cal N}_{2,3}^2)$.
The  corresponding NC couplings in Eq. (\ref{NC}) are now expressed  through 
\beq
P = {\cal N}^2.
\eeq
It is also convenient to define
\begin{equation}
{\cal N}_i^2 \equiv (1+h_i^2)^{-1}, \,\,\,\,\,\, i=1,2(3),
\label{hdef}
\end{equation}
where the $h_i$ parameters reflect the deviation from the {\sl standard}  
neutrino coupling. 

Before entering into the discussion of the resonant massless-neutrino 
conversion in matter, we describe the present upper limits from 
laboratory experiments on the relevant  parameters $h_i^2$ and $\theta$.
We first note that the laboratory limits on the leptonic
mixing angle $\theta$ are rather weak since 
no oscillations between two strictly massless neutrinos can develop
in vacuum.
However, although not strictly justified {\sl a priori} from the
point of view of laboratory constraints, we will assume 
the small-mixing angle approximation. This will be justified
 {\sl a posteriori} in view of our results. 
In this way we have $\nu_i \sim \nu_a\ [a=e,\mu (\tau)]$, and 
we can analogously interpret $h^2_i$ as  $h^2_a$. 

There have been extensive studies of experimental universality 
tests which restrict the parameters $h_a^2$. For the case of
$h^2 _\tau$ one can still allow values in the range of a few percent  
\cite{LL,univ}, whereas the constraints on $h_e^2$ and  $h^2 _{\mu}$ 
are more stringent. 
Therefore, from now on we focus on the $(\nu_e,\nu_\tau)$ system, 
for which the universality limits are the weakest.  
Moreover, the present experimental situation cannot exclude that 
the difference $h^2_{\tau} -h^2_{e}$ can be positive as  
required later on in our discussion. 

\section{Resonant Massless-Neutrino Conversion}
\

Here we briefly recall the main features of the resonant 
conversion mechanism of massless neutrinos emerging from the 
previous scenario. 
For convenience we choose to write the system of 
Schroedinger equations, which describe the propagation of 
the two neutrinos in matter, in the basis defined as \cite{valle}
\beq
\label{basis}
\tilde{\nu}_a \equiv [R {\cal N}^{-1} R^T]_{ab} \: {\nu_b}, \:\:\: a,b 
= e,\tau.
\eeq
Although this basis is somehow artificial, it almost coincides with the
flavour basis for small lepton universality violation or small mixing
angle $\theta$. In this basis, the Schroedinger equations can be written as
\begin{equation}
{i{d \over dr}\left(\matrix{
\tilde{A}_e \cr\ \tilde{A}_\tau\cr }\right)=
 \sqrt{2} G_F {\rho \over m_N}
 \left(\matrix{
 \tilde{H}_{e}
& \tilde{H}_{e\tau} \cr
 \tilde{H}_{e\tau} 
& \tilde{H}_{\tau} \cr}
\right)
\left(\matrix{
\tilde{A}_e \cr\ \tilde{A}_\tau \cr}\right) },
\label{evolution1}
\end{equation}
where $\tilde{A}_{e,\tau}$ are the amplitudes corresponding
to the neutrino states in the basis of \Eq{basis}, $G_F$ 
is the Fermi constant, $\rho$ is the matter density, and $m_N$ is 
the nucleon mass. The entries of the evolution Hamiltonian are 
now given by
\begin{eqnarray}
\label{hamilt}
\tilde{H}_{e} &  = & Y_e({\cal N}_e c^2+{\cal N}_\tau s^2)^2-
\frac{1}{2}Y_n({\cal N}_e^2c^2+{\cal N}_\tau^2s^2), 
\nonumber \\
\tilde{H}_{e\tau} & = & \tilde{H}_{\tau e}=
[Y_e({\cal N}_e c^2+{\cal N}_\tau s^2)-\frac{1}{2}Y_n({\cal N}_e
+{\cal N}_\tau)]({\cal N}_\tau-{\cal N}_e)sc, \\
\tilde{H}_{\tau} & = & Y_e s^2c^2({\cal N}_\tau^2-{\cal N}_e^2)
-\frac{1}{2}Y_n({\cal N}_e^2s^2+{\cal N}_\tau^2c^2), \nonumber
\end{eqnarray}
where for brevity we have used the short-hand notations
$s=\sin\theta$ and $c=\cos\theta$. 

In an electrically neutral medium, $Y_e$ and $Y_n$ are defined as
\begin{equation}
Y_e \equiv \frac{n_e}{n_e+n_n} \, ,\hskip 1cm Y_n = 1-Y_e\, ,
  \label{Ye}
\end{equation}
where $n_e$ and $n_n$ are the net electron and the neutron 
number densities in matter, respectively. Note that the evolution 
matrix has no energy dependence, which implies that for the 
corresponding antineutrino system ($\bar{\nu}_e, \bar{\nu}_\tau$) 
this matrix just changes its overall sign. 
Clearly, in this scenario, resonant neutrino conversion can also occur 
provided the condition $\tilde{H}_{e} = \tilde{H}_{\tau}$ 
is fulfilled \cite{valle}.
In fact, the same resonance condition holds for both 
$\nu_e\leftrightarrow \nu_\tau$ and 
$\bar{\nu}_e \leftrightarrow \bar{\nu}_\tau$ channels.
As a result, in the thermal phase of supernova neutrino emission,
{\sl both} neutrinos {\sl and} antineutrinos can simultaneously
undergo this resonance. This will be very important for our subsequent 
discussion in Sec. 4.

In order to simplify \Eq{evolution1}, we take advantage of the small
parameters $h_a^2$ expected from the 
universality constraints. With the previous assumption of small $\theta$, 
we obtain
\begin{equation}
{i{d \over dr} \left(\matrix{
\tilde{A}_e \cr\ \tilde{A}_\tau \cr }\right)=
\sqrt{2} G_F {\rho \over m_N}
 \left(\matrix{
 Y_e-\frac{1}{2} Y_n (1-h^2_e)
& {1\over 2} \eta(Y_n- Y_e )\sin2\theta \cr
 {1\over 2} \eta(Y_n- Y_e )\sin2\theta 
& -\frac{1}{2} Y_n (1-h^2_\tau) \cr}
\right)
\left(\matrix{
\tilde{A}_e \cr\ \tilde{A}_\tau \cr}\right) },
\label{evolution2}
\end{equation}
\noindent
where the parameter $\eta$ is defined as 
\beq
\eta \equiv \frac{1}{2} (h_\tau^2-h_e^2). 
\eeq
The mixing angle  $\theta_m$ and the
neutrino oscillation length $L_m$ in matter are given by
\begin{equation}
\sin^2 2\theta_m
= \frac{ \eta^2(Y_n-Y_e)^2 \sin^2 2\theta}{ (Y_e-\eta Y_n)^2 +
\eta^2(Y_n- Y_e )^2 \sin^2 2\theta },
  \label{mixinganglesin}
\end{equation}
\begin{equation}
L_m=
 \frac{2\pi}{\sqrt{2} G_F ({\rho / m_N})
[ (Y_e-\eta Y_n)^2 +
\eta^2(Y_n- Y_e )^2 \sin^2 2\theta ]^{1/2}},
  \label{length}
\end{equation}
respectively.

The resonance condition now reads
\begin{equation}
Y_e = \eta Y_n.
\label{rc2}
\end{equation}
Here we should stress that a positive value of $\eta$ is necessary for
the above equation to hold. 
Moreover, due to the bounds on the lepton universality violation, 
$\eta \lsim$ $O(10^{-2})$, the condition in \Eq{rc2} can
be fulfilled only in a 
strongly-neutronized medium. 
This is why the present mechanism cannot work in the  matter 
background of the Sun ($Y_n \leq 0.33$) \cite{valle,LL} or Earth 
($Y_n \sim 0.5$). On the other hand, the material composition 
just above the neutrinosphere in type II supernovae ($Y_e \ll Y_n$) 
can satisfy \Eq{rc2}, as shown later.

In our subsequent discussion, we will employ the simple Landau-Zener 
approximation \cite{Landau,HPD} to estimate the conversion 
probability after the neutrinos cross the resonance. Under this 
approximation, the probability for $\nu_e\leftrightarrow\nu_\tau$
and $\bar\nu_e\leftrightarrow\bar\nu_\tau$ conversions is given by
\begin{eqnarray}
\label{LZ}
  P & = & 1 - 
\exp\Biggl(-\dfrac{\pi^2}{2}\dfrac{\delta r}
{L_{m}^{\rm res}} \Biggr) \nonumber \\
          & \approx &  \exp\left[
-32 \times \left(\dfrac{\rho_{\rm res}}{10^{12} {\mbox{g/cm}^3}}\right)
      \Biggl( \dfrac{\eta}{10^{-2}}     \Biggr)
      \Biggl(\dfrac{H}{\mbox{cm}} \Biggr)\sin^22\theta 
                 \right], \nonumber \\
 \delta r & = & 2 H\sin 2\theta, \,\,\,\,\,\,\,\,
H \equiv \left| \frac{\mbox {d}\ln Y_e}{\mbox{d}r}\right|^{-1}_{\rm res}, 
\end{eqnarray}
where $L_{m}^{\rm res}$ is the neutrino oscillation length
at resonance and $\rho_{\rm res}$ are 
the corresponding matter density. In deriving the above equation, we
have used $Y_n\approx 1$ for the neutron abundance near resonance.
Notice that for $\delta r /L_{m}^{\rm res}>1$
resonant neutrino conversion will be adiabatic \cite{MS}. 
It is also important to note that the conversion 
probability does not depend on the 
neutrino energy [cf. Eqs. (\ref{evolution1}) and (\ref{hamilt})].

\vglue 1cm

\section{Massless-Neutrino Conversion in Supernovae}
\subsection{Neutrino Emission and $Y_e$ Profile in Supernovae}
\

A supernova occurs when the core of a massive star collapses into a compact
neutron star. Almost all of the gravitational binding energy of the final
neutron star is radiated in $\nu_e$, $\bar\nu_e$, $\nu_\mu$, $\bar\nu_\mu$,
$\nu_\tau$, and $\bar\nu_\tau$. The last four neutrino species are created
by thermal pair production processes inside the neutron star. On the other
hand, although most of the $\nu_e$ and $\bar\nu_e$ are
produced in pairs, there is a net excess of $\nu_e$ over $\bar\nu_e$ due to
the neutronization and deleptonization of the core through 
$e^-+p\rightarrow n+\nu_e$.
Because all these neutrinos have 
intense neutral-current scatterings on the free nucleons 
inside the neutron star, the net lepton number carried by $\nu_e$
can escape from the neutron star only through diffusion.
Therefore, we expect to see the strongest deleptonization
effect near the neutrinosphere, where neutrinos stop diffusing
and begin free-streaming. 

We can estimate the electron fraction
near the neutrinosphere as follows. 
{}From the approximate chemical equilibrium
for $e^-$, $p$, $n$, and $\nu_e$, we have 
\begin{equation}
\mu_{e^-}+\mu_p\sim \mu_n,
\end{equation}
where for example, $\mu_{e^-}$ is the electron chemical potential, and
where we have set $\mu_{\nu_e}\sim 0$. For non-relativistic nucleons,
we can write
\begin{equation}
{n_n\over n_p}\sim \exp\left({\mu_n-\mu_p\over T}\right),
\end{equation}
where $T$ is the temperature, and where we have neglected the 
neutron-proton mass difference. The electron fraction is then given by
\begin{equation}
\label{secondye}
Y_e\equiv{n_p\over n_p+n_n}\sim {1\over \exp(\mu_{e^-}/T)+1}.
\end{equation}
The chemical potential for relativistic and degenerate
electrons near the neutrinosphere is approximately given by
\begin{equation}
\label{mupot}
\mu_{e^-}\approx(3\pi^2n_e)^{1/3}\approx 51.6(Y_e\rho_{12})^{1/3}\ \mbox{MeV},
\end{equation}
where $\rho_{12}$ is the matter density in units of $10^{12}$ g cm$^{-3}$.
For typical conditions near the neutrinosphere, $T\sim 4$ MeV and 
$\rho_{12}\sim 10$, by solving Eqs. (\ref{secondye}) and (\ref{mupot}),
we find $Y_e\sim 6\times 10^{-3}$, in good agreement with
the numerical supernova models. Therefore, we 
can expect resonant massless-neutrino conversions to occur above
the neutrinosphere as long as 
the lepton non-universality parameter $\eta \gsim 6 \times 10^{-3}$
[cf. Eq. (\ref{rc2})]. 

Above the neutrinosphere, the approximate chemical equilibrium between
$\nu_e$ and matter no longer holds. The electron fraction is determined 
by the following reactions:
\begin{eqnarray}
\label{nu-n}
\nu_e+n&\rightleftharpoons&p+e^-,\\
\label{nu-p}
\bar\nu_e+p&\rightleftharpoons&n+e^+.
\end{eqnarray}
In fact, Qian {\it et al.} \cite{QFMMW} 
have shown that $Y_e$ above the neutrinosphere is given by
\begin{equation}
Y_e\approx{\lambda_{e^+n}+\lambda_{\nu_en}\over\lambda_{e^-p}+\lambda_{e^+n}
+\lambda_{\bar\nu_ep}+\lambda_{\nu_en}},
\end{equation}
where for example, $\lambda_{\nu_en}$ is the rate for the forward reaction
in Eq. (\ref{nu-n}). 
In particular, because $\lambda_{e^-p}$ and $\lambda_{e^+n}$
quickly decrease with the temperature, the asymptotic value of $Y_e$ 
at large radii is approximately given by
\begin{equation}
\label{thirdye}
Y_e\approx{\lambda_{\nu_en}\over\lambda_{\bar\nu_ep}+\lambda_{\nu_en}}.
\end{equation} 
Therefore, the asymptotic electron fraction above the neutrinosphere 
is essentially determined by the characteristics of the $\nu_e$ and
$\bar\nu_e$ fluxes, such as their luminosities and energy distributions.

The individual neutrino luminosities in supernovae are approximately the
same:
\begin{equation}
\label{lumin}
L_{\nu_e}\approx L_{\bar\nu_e}\approx L_{\nu_{\tau(\mu)}}\approx
L_{\bar\nu_{\tau(\mu)}}.
\end{equation}
However, the individual neutrino energy distributions are very different.
This is because these neutrinos have different abilities to exchange energy
with the neutron star material, and thermally decouple at different
temperatures inside the neutron star. 
Unlike $\nu_e$ and $\bar\nu_e$, $\nu_{\tau(\mu)}$ and $\bar\nu_{\tau(\mu)}$ 
are not energetic enough
to have charged-current absorptions on the free nucleons inside the
neutron star. Furthermore, between $\nu_e$ and $\bar\nu_e$, $\nu_e$ have
more frequent absorptions due to the high neutron abundance in the
neutron star matter. 
As a result, $\nu_{\tau(\mu)}$ and $\bar\nu_{\tau(\mu)}$
thermally decouple at the highest temperature, and $\nu_e$ decouple at
the lowest temperature. Correspondingly, the average neutrino energies
satisfy the following hierarchy:
\begin{equation}
\label{hierarchy}
\langle E_{\nu_e}\rangle <\langle E_{\bar\nu_e}\rangle <
\langle E_{\nu_{\tau(\mu)}}\rangle 
\approx\langle E_{\bar\nu_{\tau(\mu)}}\rangle.
\end{equation}
Typically, the average supernova neutrino energies are: 
\begin{equation}
\label{average}
\langle E_{\nu_e}\rangle \approx 11\ \mbox{MeV},\ \langle E_{\bar\nu_e}\rangle
\approx 16\ \mbox{MeV},\ \langle E_{\nu_{\tau(\mu)}}\rangle \approx \langle
E_{\bar\nu_{\tau(\mu)}}\rangle\approx 25\ \mbox{MeV}.
\end{equation} 

Now we can understand the electron fraction profile in supernovae as
illustrated in Fig. 1. In this figure, we plot the typical electron fraction
and density 
profiles in Wilson's supernova model at time $t>1$ s after the explosion. 
The solid line is for the density, and the dotted line is for $Y_e$. 
As we can see,
the minimum value of $Y_e$ occurs near the neutrinosphere.
Above the neutrinosphere, the electron fraction
is set by the reactions in Eqs. (\ref{nu-n}) and (\ref{nu-p}). 
At large radii, it reaches
an asymptotic value much larger than the minimum $Y_e$.  

\vspace{1.cm}

{}From the above discussion of neutrino emission and $Y_e$ profile in
supernovae, we find that it is
interesting to study massless-neutrino conversion in supernovae.
First of all, the resonance condition for such conversion, Eq. (\ref{rc2}) 
can be fulfilled above the neutrinosphere for $\eta\sim 0.01$. 
Furthermore, conversion
between $\nu_\tau \ (\bar\nu_\tau)$ and $\nu_e \ (\bar\nu_e)$ can alter
the supernova neutrino characteristics, especially the average neutrino
energies in Eq. (\ref{average}). 
We can gauge the potential to use supernovae as a sensitive probe of
the mixing between massless neutrinos by
estimating the adiabatic condition for resonant massless-neutrino conversion. 
For $\eta\sim 10^{-2}$, the resonances occur at densities
$\rho \sim 10^{12}$--$10^{13}$g cm$^{-3}$, just above the neutrinosphere. 
The corresponding scale height for $Y_e$ is $H\sim 1$--10 km. 
{}From Eq. (\ref{LZ}), we see that massless neutrinos can be adiabatically
converted for $\sin^2 2 \theta > 10^{-7}$--$10^{-6}$. 
In the following subsections, we discuss two possible ways to
probe the mixing between massless neutrinos in supernovae.

\subsection{Detection of $\bar\nu_e$ from SN1987a}
\

The Kamiokande II and IMB detectors observed 11 and 8 $\bar\nu_e$ events,
respectively, from SN1987a [9,10]. 
An estimate of the average supernova $\bar\nu_e$ 
energy can be made from the detection data, although the obtained estimate
should be taken with caution, considering the poor statistics and
the marginal agreement between the two sets of data. 
Nevertheless, if we adopt
the standard average neutrino energies predicted by the numerical supernova
models, then a significant amount of conversion between $\bar\nu_\tau$ and
$\bar\nu_e$ can probably be ruled out. This is because the average $\bar\nu_e$
energy inferred from the detection data is much smaller than the average
$\bar\nu_\tau$ energy predicted by the numerical supernova models.
Specifically, with $\bar\nu_e\leftrightarrow\bar\nu_\tau$ conversion,
the $\bar\nu_e$ flux at the detectors would be given by
\begin{equation}
\phi_{\bar \nu_e} =
 \phi_{\bar \nu_e}^0 (1-P)
+  \phi_{\bar{\nu}_\tau}^0 P, 
\label{spectrum}
\end{equation}
where $\phi_{\bar\nu_e}^0$ and $\phi_{\bar\nu_\tau}^0$ are the
$\bar\nu_e$ and $\bar\nu_\tau$ fluxes in the absence of
neutrino conversions, respectively,
and $P$ is the conversion probability. For large $P$, based on predictions
from numerical supernova models, the $\bar\nu_e$ energy spectra at the
detectors would have been significantly harder than detected in the case
of SN1987a. From the detection data,
Smirnov {\sl et al.} 
\cite{SSB} argued that the probability 
for $\bar\nu_e\leftrightarrow\bar\nu_{\tau(\mu)}$ conversion should be less
than 0.35.

We can apply the same argument to constrain the
mixing between massless neutrinos.
Using the density and $Y_e$ profiles from Wilson's supernova model in Fig. 1,
we plot in Fig. 2 two contours of the conversion probability
in the ($\eta,\sin^2 2 \theta$) parameter space. The solid line is for
a conversion probability of $P\approx0.5$, and the dashed line is for
$P\approx 0.35$. We can conclude that mixing between massless 
$\bar\nu_e$ ($\nu_e$) and $\bar\nu_\tau$ ($\nu_\tau$) at a level of
$\sin^2 2 \theta \gsim 10^{-6}$ is ruled out for $\eta\gsim 10^{-2}$
due to the non-observation of unexpectedly hard $\bar\nu_e$ energy
spectra from SN1987a. Such a stringent upper limit
on the mixing angle $\theta$ justifies the approximation we have made 
in deriving \Eq{evolution2}. 

\vglue 1cm

\subsection{Supernova $r$-process Nucleosynthesis}
\

Now we consider the effect of massless-neutrino conversions on the 
the supernova $r$-process nucleosynthesis. The $r$-process is responsible
for synthesizing about a half of the heavy elements 
with mass number $A>70$ in nature.
It has been proposed that the $r$-process occurs in the region above the
neutrinosphere in supernovae when significant neutrino fluxes are still
coming from the neutron star \cite{Woosley}. 
A necessary condition required for the
$r$-process is $Y_e<0.5$ in the nucleosynthesis region. As we have discussed
previously, the $Y_e$ value at large radii above the neutrinosphere, where
the $r$-process nucleosynthesis takes place, is
determined by the neutrino absorption rates
$\lambda_{\nu_en}$ and $\lambda_{\bar\nu_ep}$. 
In turn, these rates depend on the $\nu_e$ and $\bar\nu_e$ luminosities and
energy distributions. 

Qualitatively, we can argue that these rates are
proportional to the product of the neutrino luminosity and average neutrino
energy. This is because the neutrino absorption rate is given by 
\begin{equation}
\label{rates}
\lambda_{\nu N}\approx \phi_\nu\,\langle \sigma_{\nu N}\rangle
\propto {L_\nu\over\langle E_\nu\rangle}\langle E_\nu^2\rangle
\propto L_\nu\langle E_\nu\rangle \, ,
\end{equation}
where $\phi_\nu$ is the neutrino flux, $\sigma_{\nu N}\propto E_\nu^2$ is
the neutrino absorption cross section, and $\langle\ \rangle$ denotes
the averaging over the neutrino energy distribution.
Therefore, the $Y_e$ in the nucleosynthesis region is approximately given
by
\begin{equation}
\label{fourthye}
Y_e\approx {\lambda_{\nu_en}\over\lambda_{\bar\nu_ep}+\lambda_{\nu_en}}
\approx {1\over 1+\langle E_{\bar\nu_e}\rangle/\langle E_{\nu_e}\rangle}.
\end{equation}
Using the average energies in Eq. (\ref{average}), 
we obtain $Y_e\approx 0.41$, in good
agreement with the numerical supernova models.

However, in the presence of massless-neutrino conversion, average
 energies of both $\bar\nu_e$ and $\nu_e$ can be affected. The 
corresponding $Y_e$ in the nucleosynthesis region is given by
\begin{equation}
\label{34}
Y_e \approx {1\over 1+{\VEV{E_{\bar\nu_e}}_{\rm eff}}/
\VEV{E_{\nu_e}}_{\rm eff}},
\end{equation}
where 
\bea
\VEV{E_{\bar\nu_e}}_{\rm eff} \equiv \VEV{E_{\bar\nu_e}} (1-P) + 
\VEV{E_{\bar\nu_\tau}} P,	\\\nonumber
\VEV{E_{\nu_e}}_{\rm eff} \equiv \VEV{E_{\nu_e}} (1-P) + 
\VEV{E_{\nu_\tau}} P.
\eea
Due to the the {\sl simultaneous} occurrence of resonant
$\nu_e \leftrightarrow \nu_\tau$ and 
$\bar{\nu}_e \leftrightarrow \bar{\nu}_\tau$ conversions, 
there is a trend to equalize the average $\nu_e$ and $\bar\nu_e$
energies, and as a result, to increase
$Y_e$ with respect to the case with no neutrino or antineutrino conversions.
For conversion probabilities of $P\approx0.15$, 0.3, and 0.8, we obtain
$Y_e\approx 0.43,$ 0.45, and 0.49. In Fig. 3, we present the contour lines 
corresponding to 
these $Y_e$ values using the density and 
$Y_e$ profiles in Wilson's supernova model. The dotted, dashed, and solid
lines in this figure are for $Y_e\approx 0.43$, 0.45, and 0.49, respectively.

In order for any $r$-process nucleosynthesis to occur, the $Y_e$ in the
nucleosynthesis region must be less than 0.5. However, in the most recent 
$r$-process model by Woosley {\it et al.} \cite{Woosley}, many of the 
$r$-process nuclei are produced only for $Y_e<0.45$. If we take 
$Y_e<0.45$ as a criterion for a successful $r$-process, then  
mixing between $\nu_e$ $(\bar\nu_e)$ and $\nu_\tau$ $(\bar\nu_\tau)$ 
at a level of $\sin^22\theta>10^{-6}$ is excluded for $\eta\gsim 10^{-2}$. 
This excluded region is similar to the previous one from considering 
the detection of $\bar\nu_e$ from SN1987a, because the limits on the 
conversion probability are about the same in both cases.   
However, we note that if the $r$-process indeed occurs in supernovae,
then the consequent limits on the mixing between massless neutrinos
are much less dependent on the predicted average neutrino energies
than the previous limits obtained by considering the $\bar\nu_e$ energy
spectra from SN1987a. This is because the $r$-process argument relies
only on the ratio of the average neutrino energies [cf. \Eq{fourthye}]. 

\subsection{Comparison of MSW and Massless-Neutrino Conversion
Mechanisms in Supernovae}
\

It is instructive at this stage to compare the effects of 
resonant massless-neutrino conversions
with those of the standard MSW mechanism in supernovae.
To simplify this comparison, we will assume small
vacuum mixing angles ($\theta\ll 1$) in both cases. 
We first note that in the MSW scenario \cite{MS}, 
for a given sign of $\delta m^2$ (e.g., $\delta m^2>0$ for
$m_{\nu_\tau}>m_{\nu_e}$),
only one kind of resonant conversion, either 
$\nu_e\leftrightarrow\nu_\tau$ (for $\delta m^2>0$), or 
$\bar{\nu}_e\leftrightarrow\bar{\nu}_\tau$ (for $\delta m^2<0$), can occur.
If $\delta m^2>0$ the MSW mechanism would not alter the $\bar\nu_e$ energy 
spectra from SN1987a, and therefore,  
no constraints on neutrino masses and mixings can be obtained for 
this mechanism from the detection data (assuming all the events were due
to $\bar\nu_e$). In contrast, severe constraints on massive-neutrino mixing
can be obtained in this
case by requiring
$Y_e<0.5$ in the nucleosynthesis region to allow 
a successful $r$-process \cite{QFMMW}.
On the other hand, if $\delta m^2<0$ the MSW mechanism could 
significantly modify the $\bar{\nu}_e$ energy spectra and
generate an excess of energetic $\bar\nu_e$. As a 
result, the parameter region which would give large probabilities
for $\bar\nu_e\leftrightarrow\bar\nu_\tau$ conversion can be possibly 
excluded by combining the predicted average supernova neutrino energies
and the SN1987a detection data.
On the contrary, significant $\bar\nu_e\leftrightarrow\bar\nu_\tau$
conversion would tend to decrease $Y_e$ in the nucleosynthesis region
(see \eq{34}) and therefore, would not conflict with the supernova 
$r$-process nucleosynthesis scenario [25].
As we can see, one can only use 
either the SN1987a detection data (for $\delta m^2<0$), or the supernova
$r$-process 
nucleosynthesis (for $\delta m^2>0$) to constrain neutrino masses and mixings
in the MSW mechanism. 

In contrast, in the case of massless-neutrino conversions, we have seen
that for $\eta\gsim 10^{-2}$, both $\nu_e\leftrightarrow\nu_\tau$
and $\bar\nu_e\leftrightarrow\bar\nu_\tau$ conversions can occur
in supernovae.
Therefore, both the SN1987a detection data and the supernova 
$r$-process nucleosynthesis should be considered in order to 
constrain the mixing of massless neutrinos.
Of course, if $\eta<0$ or $\eta\ll 10^{-2}$, then no resonant 
massless neutrino conversions would occur in supernovae.
The constraints on massless-neutrino mixing in this case 
are perhaps hard to obtain by any means.

It is interesting to note that simultaneous $\nu_e\leftrightarrow
\nu_\tau$ and $\bar\nu_e\leftrightarrow\bar\nu_\tau$ conversions
would give rise to distinctive supernova neutrino signals in 
large volume detectors, such as super-Kamiokande [26], SNO [27],
and LVD [28]. For example, in the super-Kamiokande detector,
the energy distributions for
both the isotropic $\bar\nu_e$ events and the forward-peaked
$\nu_e$ events would be altered.
With enough statistics, such detectors may be able
to distinguish the massless-neutrino conversion scenario from
the standard MSW mechanism, should any neutrino conversion 
indeed occur in supernovae.
%

\vglue 0.5cm
\section{Conclusions}
\vglue 0.5cm

We have discussed the possibility of probing the mixing  
between massless neutrinos described in the theoretical scheme 
in Sec. 2. Due to the relatively stringent laboratory bounds on the
weak universality violation, the supernova matter background 
seems to be the unique site where resonant conversions of  
massless neutrinos can take place. 
By considering the detection of $\bar\nu_e$ from SN1987a and the 
supernova $r$-process nucleosynthesis, we have obtained 
stringent limits on the mixing between massless $\nu_e$ ($\bar\nu_e$)
and $\nu_\tau$ ($\bar\nu_\tau$) presented in Figs. 2 and 3.
These limits, at a level of $\sin^22\theta\lsim 10^{-6}$, are
rather remarkable, because the usual laboratory methods
to constrain neutrino mixing through vacuum neutrino oscillation 
searches are totally insensitive to the mixing between massless neutrinos. 
Indeed, the supernova limits we have obtained for the mixing between 
massless $\nu_e$ ($\bar\nu_e$) and $\nu_\tau$ ($\bar\nu_\tau$) are
orders of magnitude more stringent than the typical 
limits on massive-neutrino mixing from laboratory neutrino
oscillation searches. 

Finally, we hope that our discussions of massless-neutrino conversions
in supernovae 
serve to highlight the interest in sharpening the 
laboratory limits on universality violation and/or pinning down 
more accurate supernova models.

\vglue 1cm
\centerline{\bf Acknowledgement}

We thank J. R. Wilson for providing us with his numerical supernova models.
We also want to thank Hideyuki Suzuki for a helpful discussion. 
This work was supported by DGICYT under Grant No. PB92-0084, 
by the Human Capital and Mobility 
Program under Grant No. ERBCHBI CT-941592 (A. R.), 
and by a DGICYT postdoctoral fellowship (H. N.).
Y.-Z. Qian was supported by a fellowship grant at Caltech.

\noindent

\newpage
\bef
\centerline{\protect\hbox{
\psfig{file=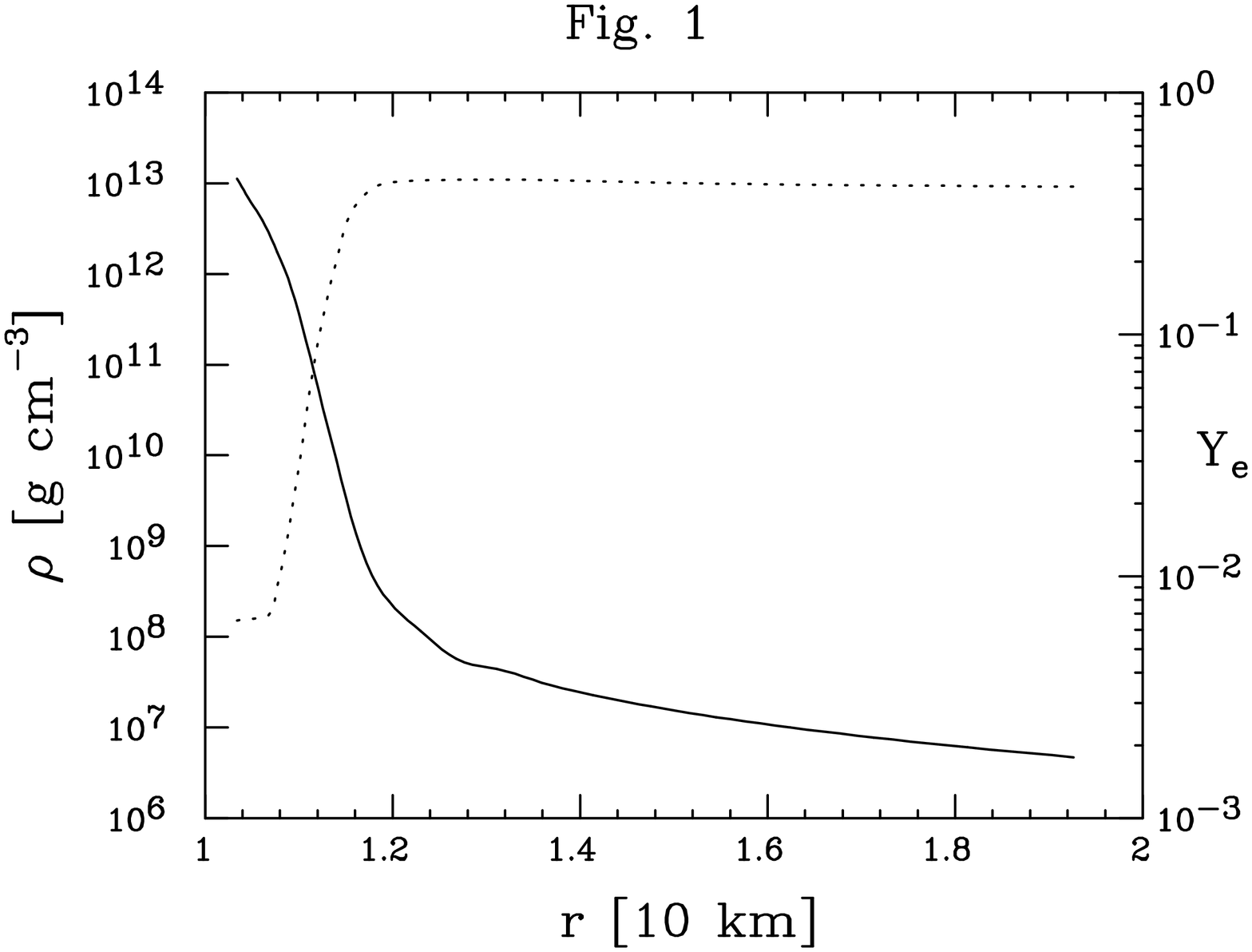,height=13.0cm,width=16.0cm,angle=90}
}}
\caption{Typical matter density (solid line) and $Y_e$ (dotted line) profiles 
in Wilson's numerical supernova model at $t>1$ s after the explosion.}
\label{0}
\eef
\bef
\centerline{\protect\hbox{
\psfig{file=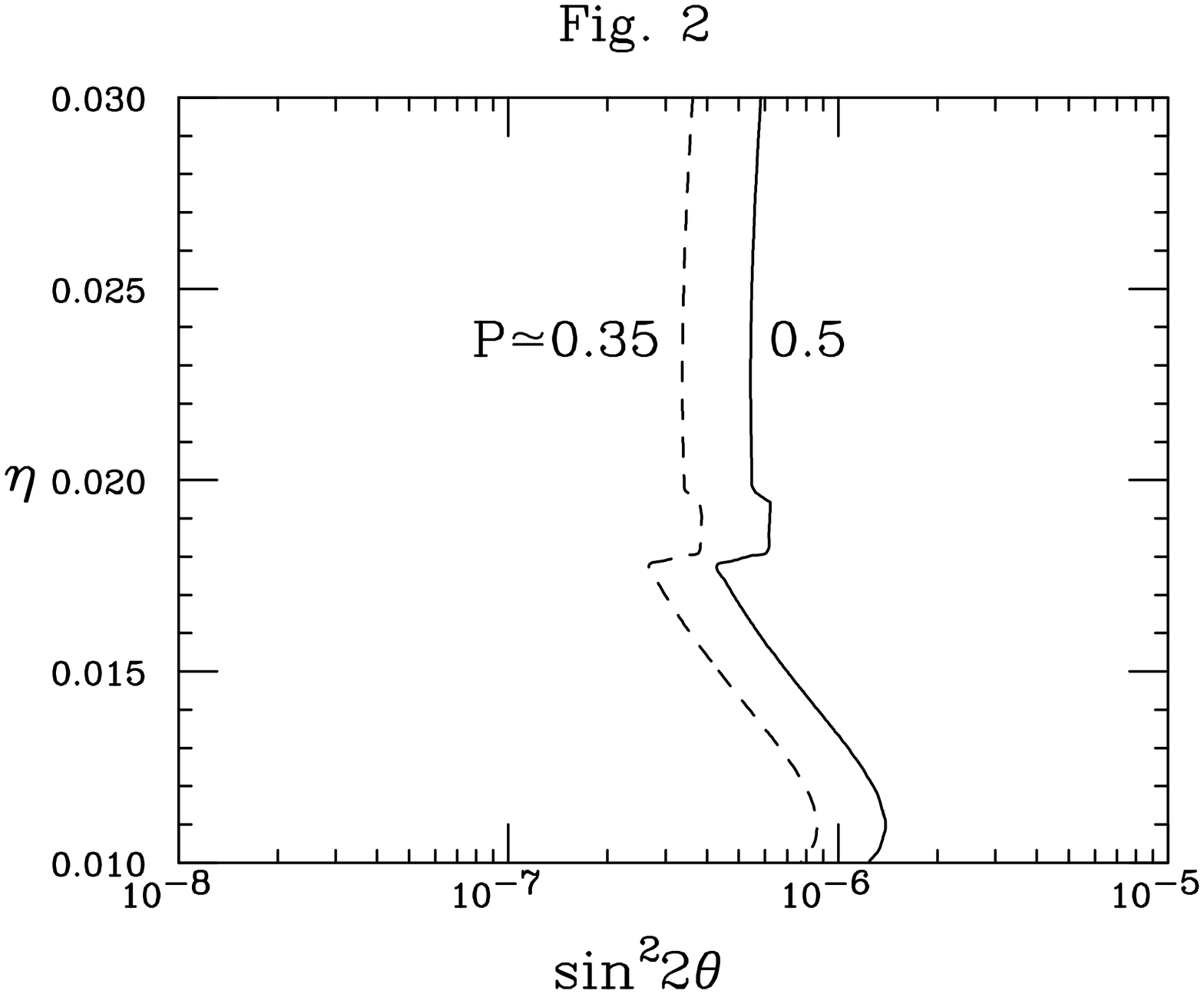,height=13.0cm,width=16.0cm,angle=90}
}}
\caption{
Constraints on massless-neutrino mixing 
from the detected SN1987a $\bar\nu_e$ energy spectra. The region to the
right of the dashed (solid) lines are excluded by the detection data
for an allowed
conversion probability of $P<0.35$ (0.5).}
\eef
\bef
\centerline{\protect\hbox{
\psfig{file=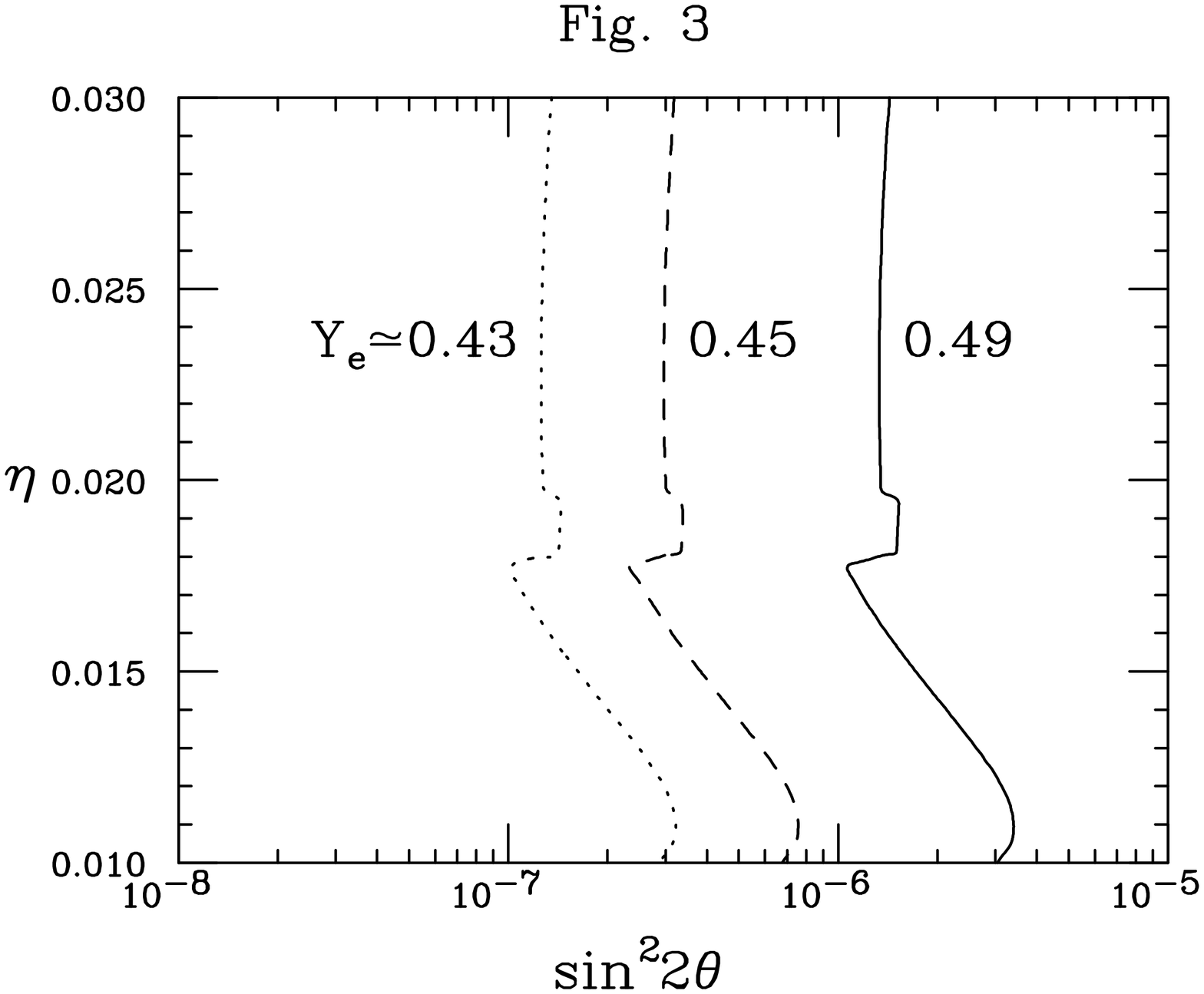,height=13.0cm,width=16.0cm,angle=90}
}}
\caption{Constraints on massless-neutrino mixing 
from the supernova $r$-process nucleosynthesis. The region to the
right of the dotted, dashed and solid lines are exclued for the 
required values of $Y_e<0.43$, 0.45, and 0.49, respectively,
in the $r$-process. }
\eef

\begin{thebibliography}{99}
%
\bibitem{MS}
S. P. Mikheyev and A. Yu. Smirnov, \sjnp{42}{913}{85}. 
\bibitem{W} 
L. Wolfenstein,   \pr{D17}{2369}{78}. 
%
\bibitem{fae}
For recent reviews see e.g.,
J. W. F. Valle, {\sl Gauge Theories and the Physics of 
Neutrino Mass}, \ppnp{26}{91}{91-171} (ed. A. Faessler), 
and G. Gelmini and S. Roulet, UCLA/94/TEP/36 and references therein.
%
\bibitem{2227}
J. Schechter and J. W. F. Valle, \pr{D22}{2227}{80}. 
%
\bibitem{NHL} 
J. Bernabeu, A. Santamaria, J. Vidal, A. Mendez and J. W. F. Valle, 
\pl{B187}{303}{87}; 
G. C. Branco, M. N. Rebelo and J. W. F. Valle, \pl{B225}{385}{89}; 
N. Rius, J. W. F. Valle, \pl{B246}{249}{90}.
%
\bibitem{MV}
R. Mohapatra and J. W. F. Valle, 
\pr{D34}{1642}{86};\\ 
I. Antoniadis {\it et al.,} \pl{B208}{209}{88}.
%
\bibitem{WW}       
D. Wyler and L. Wolfenstein, \np{B218}{205}{83}.
%
\bibitem{valle}
J. W. F. Valle,  \pl{B199}{432}{87}.
%
\bibitem{LL}
P. Langacker and D. London, \pr{D38}{907}{88}.
%
\bibitem{KP}
See for a review e.g.,  
{\it Neutrinos in Physics and Astrophysics}, C. W. Kim and A. Pevsner, 
Harwood Academic Publishers, 1993.
\bibitem{BPW}
V. Barger, R. J. N. Phillips and K. Whisnant\pr{D44}{1629}{91}.
%
\bibitem{BAH}
See for e.g. J. N. Bahcall, {\it Neutrino Astrophysics}, Cambridge University 
Press 1989. 
%
\bibitem{KA}
K. Hirata {\it et. al.}, \prl{58}{1490}{87}.
%
\bibitem{IMB}
R. Bionta {\it et. al.}, \prl{58}{1494}{87}.
%
\bibitem{old}
P. Reinartz and L. Stodolsky,  \Zp{C27}{507}{85};\\
L. Wolfenstein,  \pl{B194}{197}{87}.
%
\bibitem{SSB}
A. Yu. Smirnov,  D. N. Spergel and J. N. Bahcall, 
\pr{D49}{1389}{94}.
%
\bibitem{JNR} 
B. Jegerlehner, F. Neubig and G. Raffelt, preprint MPI-PTh 95-120, 
astro-ph/9601111.
%
\bibitem{QFMMW}
Y.-Z. Qian {\it et al.,} 
\prl{71}{1965}{93}.
%
\bibitem{CKM}
M. Kobayashi and T. Maskawa, 
\ptp{49}{652}{73}.
%
\bibitem{TH}
Note that non-perturbatively, only the anomaly-free combinations $B-L$,
$L_e-L_\mu$, $L_e-L_\tau$, or $L_\mu-L_\tau$ are conserved, where
$B$ is the baryon number and $L=L_e+L_\mu+L_\tau$. See
G. 't Hooft,  \prl{37}{8}{76}; \pr{D14}{3422}{76}.
%
\bibitem{Witten} 
E. Witten, \np{B268}{79}{86}.
%
\bibitem{beyond}
For a recent review see e.g., J. W. F. Valle, in
{\sl Physics Beyond the Standard Model}, lectures given at the 
{\sl VIII Jorge Andre Swieca Summer School} (Rio de Janeiro, February 1995) 
and at {\sl V Taller Latinoamericano de Fenomenologia de las Interacciones
Fundamentales} (Puebla, Mexico, October 1995); hep-ph/9603307. 
%
\bibitem{univ}
M. Gronau, C. N. Leung and J. L. Rosner, \pr{D29}{2539}{84};
P. Langacker and D. London, \pr{D38}{886}{88};
A. Ilakovac and A. Pilaftsis, \np{B437}{491}{95};
E. Nardi, E. Roulet and D. Tommasini, \pl{B344}{225}{95}. 
%
\bibitem{Landau} 
L. Landau,  Phys. Z. Sowjetunion {\bf 2}, 46 (1932);\\
C. Zener,  Proc. R. Soc. London {\bf A137}, 696 (1932).
%
\bibitem{HPD}
W. C. Haxton, 
 Phys. Rev. Lett. {\bf 57}, 1271 (1986); \\
%
S. J. Parke, 
Phys. Rev. Lett. {\bf 57}, 1275 (1986); \\
A. Dar {et al.},  Phys. Rev. {\bf D35}, 3607 (1987).  
%
\bibitem{Woosley}
S. E. Woosley and E. Baron,  Astrophys. J. {\bf 391}, 228 (1992); \\
S. E. Woosley,  Astron. Astrophys. Suppl. Ser. {\bf 97}, 205 (1993);\\
S. E. Woosley and R. D. Hoffman,  Astrophys. J. {\bf 395}, 202 (1992);\\
B. S. Meyer {\it et al.,} Astrophys. J.  {\bf 399}, 656 (1992); \\
S. E. Woosley {\it et al.}, Astrophys. J. {\bf 433}, 229 (1994).
%
\bibitem{QF}
Y.-Z. Qian and G. M. Fuller, Phys. Rev. {\bf D52}, 656 (1995). 
%
\bibitem{SK}
Y. Totsuka, {\it Proceedings of International Symposium on 
Underground Physics Experiments}, edited by K. Nakamura, 
Institute for Cosmic Ray Research, Tokyo, 1990. 
%
\bibitem{SNO}
SNO Collaboration Proposal, SNO-87-12.
%
\bibitem{LVD}
Alberini {\it et al.,}  
LVD Collaboration,  Il Nuvo Ciment {\bf 9C}, 237 (1986).
%
\end{thebibliography}
\end{document}